\documentclass{elsart}
\usepackage{natbib}
\usepackage{psfig}
\usepackage{rotate}
\usepackage{epsfig}
\usepackage{graphicx}
\usepackage{amsfonts}
\usepackage{amssymb}

\def\ST{{\tt Slav\-nov-Ta\-y\-lor }}
\def\MT{{\tt Mathematica }}
\def\code#1{\large{\tt #1}\normalsize}
\def\cal{\rm }
\newcommand\desc[2]{{\em #1}: {#2}\\}
\begin{document}
\begin{frontmatter}
%
%
\title{{\tt Slavnov-Taylor}{\tt 1.0}: A Mathematica package for computation in BRST formalism}
\author[INFN,Murcia]{Marco Picariello}
\author[Murcia,CERN]{Emilio Torrente-Lujan}
\address[INFN]{INFN sezione di Milano, Via Celoria 16, I20133 Milano, Italy}
\address[Murcia]{Departamento de Fisica, Grupo de Fisica Teorica,
Universidad de Murcia, Murcia, Spain}
\address[CERN]{CERN-TH, CH-1211 Geneve 23, Switzerland}
\begin{abstract}
\ST is a Mathematica package which allows us to
perform automatic symbolic computation in BRST formalism.
This article serves as a self-contained guide to prospective users, and
indicates the conventions and approximations used.
\end{abstract}
\end{frontmatter}

\section{Program summary}

\desc{Title of program}{\ST}
\desc{Available at}{http://pcpicariello.mi.infn.it/ST/}
\desc{Programming Language}{Mathematica 4.0}
\desc{Platform}{Any platform supporting Mathematica 4.0}
\desc{Computers tested on}{Pentium PC}
\desc{Operating systems under which the program has been tested}{Linux}
\desc{Memory required to execute}{Minimal: 1.254.784 bytes, Standard: 1.281.248 bytes}
\desc{No. of bytes in distributed program, including test data,
etc}{35.636 bytes}
\desc{Keywords}{Slavnov-Taylor, BRST, Mathematica}
\desc{Nature of physical problem}{Symbolic computation in the
Slavnov-Taylor formalism for gauge theories in 4-Dimensional space-time
based on a semi-simple compact Lie group for a general BRS
transformations.}
\desc{Restrictions on the complexity of the problem}{Only matter in
the adjoint is allowed.}
\desc{Typical running time}{less than one second}

\section{Introduction}
Symbolic computation in the Slavnov-Taylor (ST) formalism is very
useful to study properties of a field theory such as quantum stability.
The main computation problem is to find the most general counter-term
compatible with the ST identity, and then absorb this
counter-term by a field and coupling constant redefinition.
This can be a not easy task if the BRS transformation is not
nilpotent~\cite{book}.
It is therefore desirable to construct a calculational tool which may 
perform the computation automatically.
In this article we present a tool (\ST) which partially solves this
problem. \ST allows us to perform generic computation of BRST
variation and construct generic polynomials of a given dimension and
ghost number. It can be used to check hand computation for generic
ST operator which may not be nilpotent. Moreover the symbolic
manipulation allows us to perform computation in general gauges.

We applied \ST to verify the correctness of the computation used to
prove the quantum stability of the Curci-Ferrari gauge in a
gluon-ghost condensation scenario~\cite{io}, where the BRS is not
nilpotent.

\subsection{The package}

\ST has been written as a \MT package. The interface with the user is
given by the Mathematica application, which can be run either in
on-line command or with the front-end feature.

The algorithm is not optimized, because the running time is not an
important point  due to the fact that people will need to run it few
times (or even once) for a given model. 

The imposition of the ST identity turns out to be equivalent to the
solution of a linear problem.
\ST is a tool whose output could be used for constructing such linear
system.

It may also be used as help for more theoretical studies such as
the stability of gauge theory when the ST is not nilpotent.

\subsection{Aims and Contents}
The main aims of this article are to provide a manual for the use of \ST,
to describe the limit of application and the notation used (to allow for
user generalization).

The rest of this paper proceeds as follows: 
\begin{itemize}
\item
The relevant parameters are presented in sec.~\ref{sec:notation}.
\item
The approximations employed are noted in sec.~\ref{sec:calculation}.
The algorithm of the calculation is also outlined.
\item
A description of how to use the package is given in
sec.~\ref{sec:run}, including information on the input.
More technical informations related to running and extending the
package are placed in appendices.
\item
Sample output from one run is displayed and explained in
appendix~\ref{sec:output}.
\item
Appendix~\ref{sec:prog} gives a more technical example of the package.
\item
Finally, in appendix~\ref{sec:objects}, a description of the relevant
modules and objects and their relation to each other is presented. 
\end{itemize}

\section{The creation of a model\label{sec:notation}}
In this section, we introduce the objects and parameters in the
\ST conventions. Translation to the actual name used in the \ST
package are shown in Appendix~\ref{sec:objects}.

\subsection{Field content}
The gauge theories in 4-Dimensional space-time considered in the
\ST package are based on a semi-simple compact Lie group. 
In a general gauge model the gauge fields are Lie algebra
valued and the belong to the adjoint representation:
$$
A_\mu(x) = A_\mu^a(x)\tau_a
$$
where the matrices $\tau_a$ are the generators of the group and obey
$$
[\tau_a,\tau_b]={\rm i} f_{abc} \tau_c,\quad\quad Tr\,\tau_a\tau_b = \delta_{ab}
$$
Matter fields can be included if they are in the adjoint
representation (which is for example the case of the gaugino for pure
Yang-Mills theory in Super-symmetry).
\\
The action is constructed as the most general gauge invariant and
power-counting renormalizable action.
We define the curvature:
$$
F_{\mu\nu}=\partial_\mu A_\nu - \partial_\nu A_\mu - i [A_\mu,A_\nu]
$$
and the invariant is given by:
$$
S_{inv} = -\frac{1}{4g^2} \int d^4x\ F_{\mu\nu} {\bf .} F^{\mu\nu}
$$
The operator ${\bf .}$ is the trace over the gauge group index.

The gauge fixing is done in the BRS way: one introduce the
Faddeev-Popov ghost and antighost fields $c(x)^a$ and $\bar c(x)^a$,
and the Lautrup-Nakanishi field $b(x)$.
The latter normally play the role of a Lagrange multiplier for the
gauge condition.
These three fields are Lie-algebra valued: $c(x)=c^a(x)\tau^a$, etc.
They are used to write the gauge fixing term, needed to extend to
quantum level the theory.
\\
Renormalization requires the introduction of an external source
coupled to the BRS variation of the fields.
This is done by adding to the action the terms
$$
S_{ext}=\int d^4x\ \sum_{\Phi} X[\Phi] {\bf .}( s \Phi) 
$$
where the sum is over the fields which have a non linear BRS
transformation. We used the \ST notation $X[\Phi]$ as the external field
coupled to the non linear BRS variation of the $\Phi$ field.

\subsection{BRST transformation}
The BRS transformation are defined as:
\begin{eqnarray}
s A_\mu &=& \partial_\mu c + {\rm i} A_\mu \wedge c\nonumber\\
s c     &=& i \frac{1}{2}c\wedge c\nonumber\\
s \bar c&=& b\nonumber\\
s b     &=& 0\nonumber
\end{eqnarray}
where the $\wedge$ operator is the external product and is defined on
two general quantities Lie algebra valued $\chi$ and $\eta$ as:
$$
(\chi\wedge \eta)^a = f_{abc}\chi^b\eta^c
$$
BRST invariance can thus be expressed by a functional identity, the
ST identity:
$$
{\cal S}(S) = 0
$$
where the nonlinear ST operator is given, for any
functional ${\cal F}$ by:
$$
{\cal S}({\cal F})=\int d^4x\ \sum_{\Phi}
\frac{\delta{\cal F}}{\delta X[\Phi]}\frac{\delta{\cal F}}{\delta\Phi} + s \phi
\frac{\delta{\cal F}}{\delta\phi} 
$$
where the sum over $\Phi$ is a over the fields with non linear BRS
transformation, while the sum over $\phi$ is over the fields with
linear BRS transformation.
\\
In case the BRS is nilpotent the gauge fixing action is given by
an $s$ variation of an integrated local polynomial:
$$
S_{gf} = s \int d^4x\ P(fields)
$$
otherwise it is a ST invariant integrated local polynomial.
\\
ST formalism allows us to study stability and unitarity of the model,
but these properties are not investigated by our package. These points
are reserved for a future version of the \ST package.

\subsubsection{Fields Properties}
The field properties used by the \ST package are:
\paragraph{Lorentz properties:}
The boson gauge fields $A^a_\mu(x)$ have a Lorentz index, while the
$b^a(x)$, $c(x)$ and $\bar c(x)$ have no Lorentz index (they are scalar).
\paragraph{Statistic:}
The boson gauge fields $A^a_\mu(x)$ and the Lautrup-Nakanishi fields
$b^a(x)$ are commuting fields, while the components of $c(x)$ and $\bar
c(x)$ are anti-commuting.
\paragraph{Canonical dimension:}
The boson gauge fields $A^a_\mu(x)$, the ghosts $c^a(x)$ and the
antighost $\bar c^a(x)$ have canonical dimension one,
while the fields $b^a(x)$ have canonical dimension two.
\paragraph{Ghost number:}
The boson gauge fields $A^a_\mu(x)$ and the Lautrup-Nakanishi fields
$b^a(x)$ are fields with ghost number zero, while the components of
$c(x)$ have ghost number one and the components of $\bar c(x)$ have
ghost number minus one.

\section{Calculation \label{sec:calculation}}
\begin{figure}{\label{fig:algorithm}
\begin{picture}(323,240)
\put(10,14){\makebox(280,10)[c]{\fbox{and the general
polynomial of given dimension and ghost number}}}
\put(10,40){\makebox(280,10)[c]{\fbox{User obtains
ST transformations of a general integrated polynomial}}}
\put(150,78){\vector(0,-1){23}}
\put(10,80){\makebox(280,10)[c]{\fbox{Package defines external action}}}
\put(150,118){\vector(0,-1){23}}
\put(10,120){\makebox(280,10)[c]{\fbox{Package defines the Anti-fields}}}
\put(150,158){\vector(0,-1){23}}
\put(10,160){\makebox(280,10)[c]{\fbox{User defines the BRS transformations}}}
\put(150,198){\vector(0,-1){23}}
\put(10,200){\makebox(280,10)[c]{\fbox{User defines the Fields}}}
\put(150,238){\vector(0,-1){23}}
\put(10,240){\makebox(280,10)[c]{\fbox{User loads the \ST package}}}
\end{picture}
\caption{Algorithm used to calculate ST transformations of general quantities in a given model. Each step (represented by a box) is detailed in the text.}}
\end{figure}
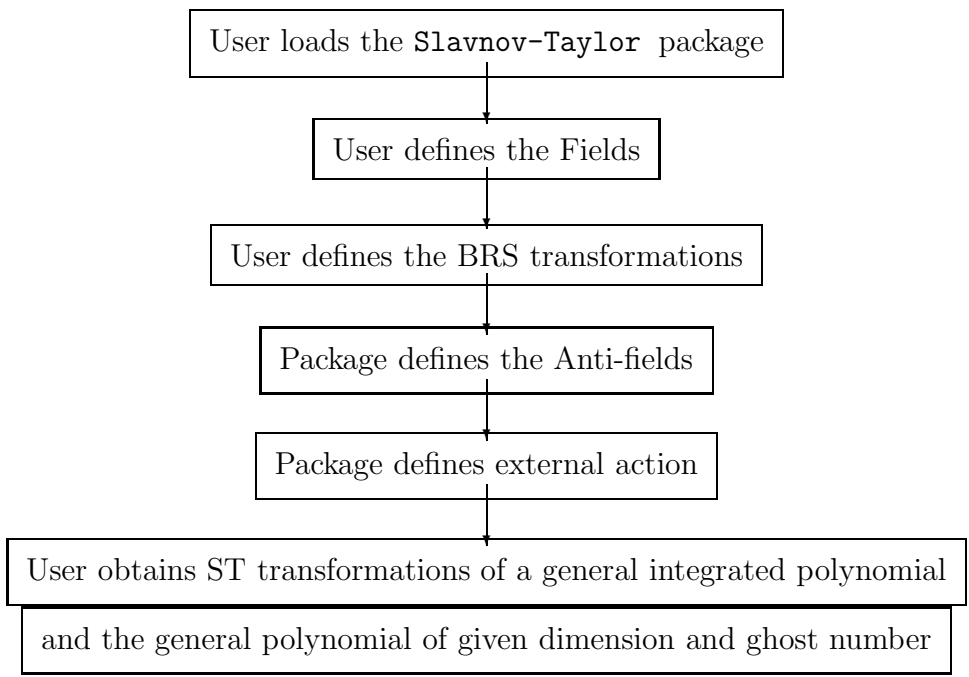

In fig~\ref{fig:algorithm}
 we show the algorithm used to perform the calculation.
Due to the fact that user can define the field content he needs and their BRS
transformation, he can easily modify the model.
For example to study pure Supersymmetric Yang-Mills theory, he needs
to define gaugino fields and the Supersymmetric BRS transformations.
In the same way the generalized BRS transformations introduced
in~\cite{Federbush:1999dw} can be treated.
Moreover theories with non nilpotent BRS transformation can be taken
into account with no problem (for example on-shell formulation of
Supersymmetric theory~\cite{susy}, or gauge theory in presence of
condensates~\cite{io2}).

\section{Running \ST}\label{sec:run}

The program \code{ST.nb} is included in the \ST distribution.
This program is can be run by Mathematica.
The program contains:
\begin{itemize}
\item{the definition of the fields $a, c, \bar c, b$
and their BRS transformation;}
\item{the invariant action and the gauge fixing Curci-Ferrari action
as an $s$ variation of a polynomial;}
\item{The definition of the full action as the sum of the invariant
action, the gauge-fixing action and the external action.}
\end{itemize}
The program gives as output:
\begin{itemize}
\item{the external action $S_{ext}$ create by the program;}
\item{the list of the field defined by the user;}
\item{the list of the external field defined by the program;}
\item{the explicit form of the full action;}
\item{the dimension of the full action;}
\item{the ghost number of the full action;}
\item{the values of $c . \frac{\delta S_{tot}}{\delta b}$}
\item{the $s$ variation of the full action (which is zero);}
\item{the most general polynomial of dimension three and ghost number
$-1$;}
\item{the most general polynomial of dimension two and ghost number
$0$;}
\item{the $s$ variation of $a_\mu . a_\mu$;}
\item{the $s$ variation of the $s$ variation of $a_\mu . a_\mu$;}
\end{itemize}

\subsection{Description of the main procedures and global parameters}

Setting the global parameter \code{\$QUIET} to a non-zero value gives
additional information on field creations.
\\
The properties of the fields content of the model are encoded in
objects and must be provided. The same for the BRS transformations.
We give here a description of the main procedures that user can use.
\begin{itemize}
 \item{\code{DefineField}
  \begin{itemize}
   \item{Parameters:
    \begin{itemize}
     \item{\code{Name} is the name of the field;}
     \item{\code{Indices} is a list which determines the Lorentz
properties of the field (an empty list define a scalar);}
     \item{\code{Type} determines the statistic of the field,
\code{Type}=\code{\$BOSE} define a bosonic (commuting) field;
\code{Type}=\code{\$FERMI} define a fermionic (anti-commuting) field;}
     \item{\code{Dimension} is the dimension of the field;}
     \item{\code{Gh} define the ghost number of the field.}
    \end{itemize}
   }
   \item{Objected created:
    \begin{itemize}
     \item{\code{Field}}
    \end{itemize}
   }
  \item{Set quantities:
   \begin{itemize}
    \item{\code{Stat} the statistic of the field: $+/- 1$ for
commuting/anti-commuting field;}
    \item{\code{Dim} the canonical dimension of the field}
    \item{\code{Gh} the ghost number of the field}
   \end{itemize}
  }
  \item{Global variables modified:
   \begin{itemize}
    \item{appends to the variable \code{\$FieldList} the name of the
  field}
   \end{itemize}
  }
 \end{itemize}
}
 \item{\code{SetBRST}
  \begin{itemize}
   \item{Parameters:
    \begin{itemize}
     \item{\code{Name} The field name of which we are defining the BRS
transformation;}
     \item{\code{brs} The BRS transformation of the field.}
    \end{itemize}
   }
   \item{Object created: 
    \begin{itemize}
     \item{\code{\$ExtField} (if the BRS transformation is not linear in
  the fields);}
    \end{itemize}
   }
  \item{Set quantities:
   \begin{itemize}
    \item{\code{s} is set to be the BRS transformation;}
    \item{\code{BRST} is set to TRUE;}
   \end{itemize}
   }
  \item{Global variable modified:
   \begin{itemize}
    \item{for non linear BRS transformations it adds to \code{Sext} the right
quantity}
   \end{itemize}
  }
  \item{Module called:
   \begin{itemize}
    \item{for non linear BRS transformations it call the module
  \code{DefineExtField} (which appends to the variable
  \code{\$ExtFieldList} the name of the external field, and works in a
  similar way than \code{DefineField} module)}
    \end{itemize}
  }
  \end{itemize}
 }
 \item{\code{Create}
  \begin{itemize}
   \item{Parameters:
    \begin{itemize}
     \item{\code{Dimension} the canonical dimension of the integrated
  polynomial we are looking for}
     \item{\code{Gh} the ghost number of the integrated
  polynomial we are looking for}
    \end{itemize}
    }
   \end{itemize}
  }
\end{itemize}
An object \code{Field} can be removed by the module
\code{UndefineField}. An object \code{ExtField} can be deleted by the
module \code{UndefineExtField}.

\appendix

\section{Sample Output \label{sec:output}}
We now present the output for a single example calculation.
When you load the package you obtain some information on the version
in use.
We load the gauge theory model with usually BRS
transformation which is defined in the \code{ST.nb} file.
Once the model is defined, we can perform the computation.
We run the commands
\small\begin{verbatim}
In[1]:= <<PackageST.m;
In[2]:= <<ST.nb;
In[3]:=
  s[a[mu].a[mu]]
  s[%]
\end{verbatim}\normalsize
The output obtained is
\small\begin{verbatim}
            mu    mu
Out[3]= -2 a   . d   c
Out[4]= 0
\end{verbatim}\normalsize
The first output is the consequence of the application of the ST operator to
quantity $a^\mu_a T_a * a^\mu_b T_b$.
The second output is the consequence of the application of the ST
operator twice to the same quantity. In this case, due to the
nil-potency of the ST operator, the result is zero.

\section{Sample Program \label{sec:prog}}
\begin{figure}[h]\label{fig:sample}
\vskip -3truecm
\hskip -3truecm
\rotatebox{90}{
\psfig{file=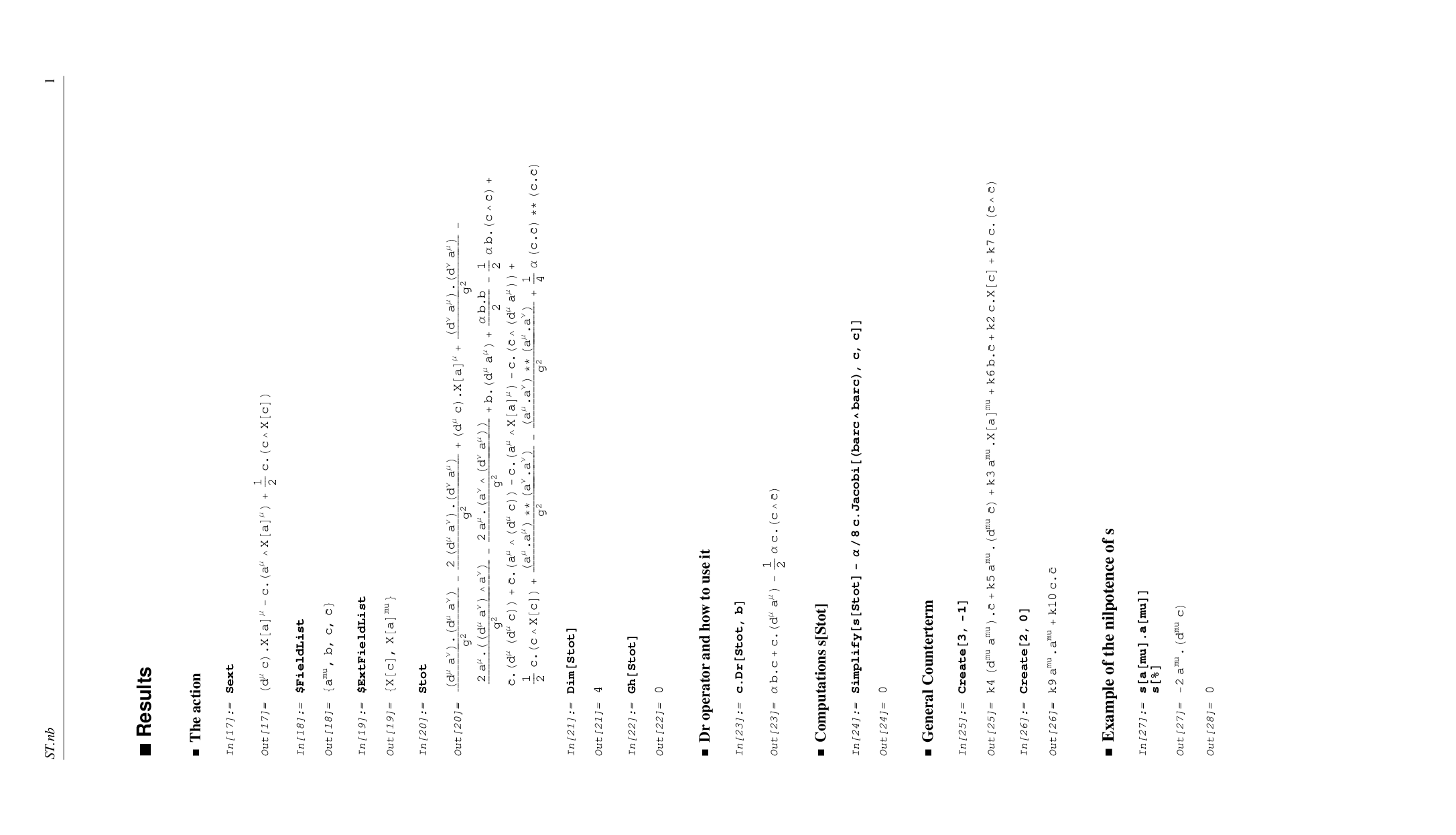,width=14truecm}}

\vskip -2truecm
\hskip -3truecm
\rotatebox{90}{
\psfig{file=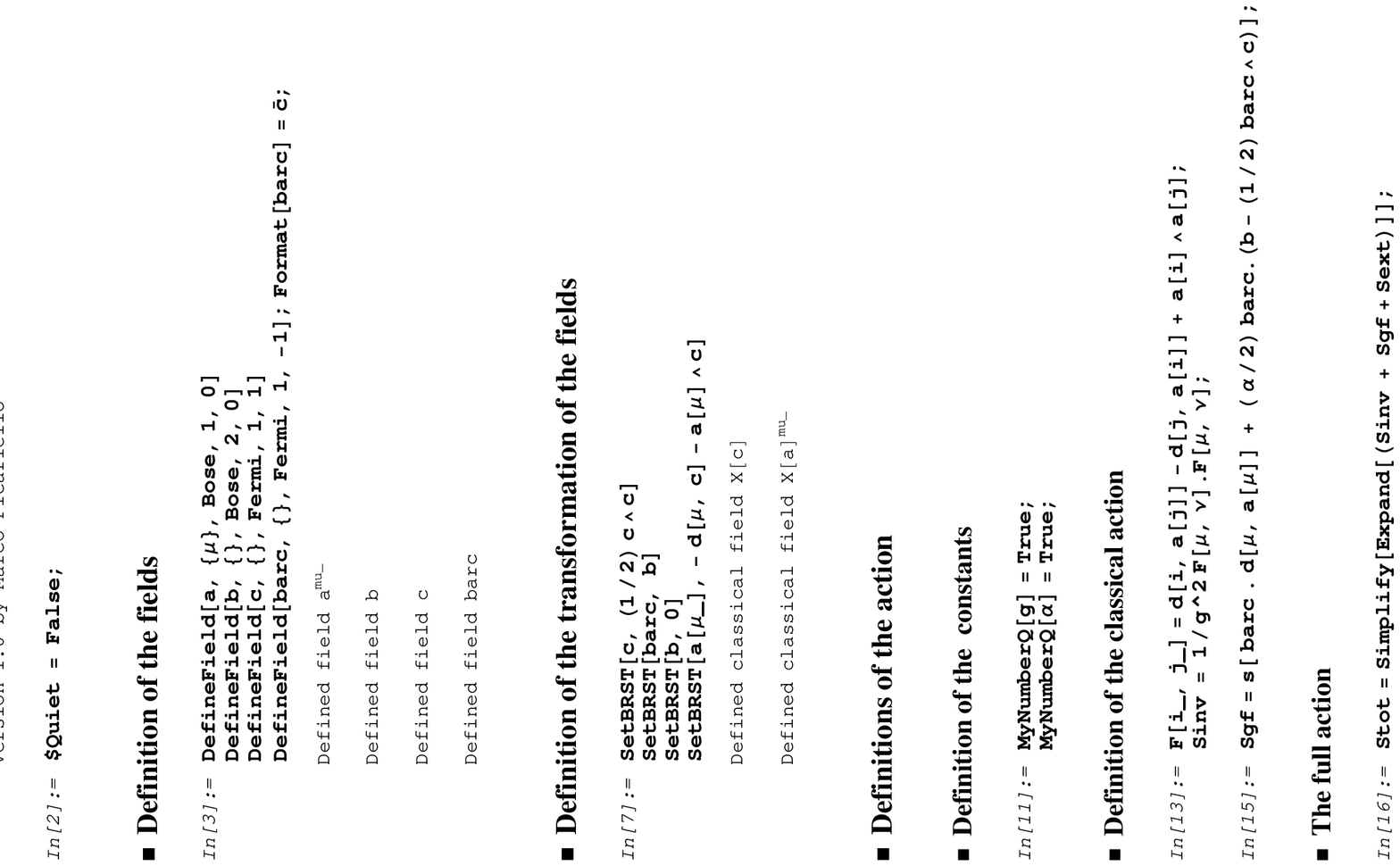,width=14truecm}}
\vskip -0.4truecm
\caption{The sample program}
\end{figure}

We now present the sample program from which it is possible to run
\ST in a simple fashion. The most important features of the modules
and objects are described in appendix~\ref{sec:objects}.
The sample program has the form displayed in figure~\ref{fig:sample}. 

After an initial introductory print-out, the fields $a,b,c,\bar c$ are
defined and their BRST transformation are set.
For these, the same notation as appendix~\ref{sec:run} is used.
\\
\begin{table}{
\caption{Detailing arguments in order for module used to define a model.}
\begin{tabular}[h]{|l|l|l|}
\hline Name & Arguments \\
\hline \code{DefineField} & Name, Lorentz indices, Statistic, Dimension,
Ghost Number\\
\code{SetBRST}	   & Name, BRST transformation
\\\hline
\end{tabular}
}\end{table}
\\
The user has created the field content of his model.
Now is time to define the classical action
and to fix the gauge (for example the Curci-Ferrari gauge):
\\
The first results that can be easily obtained are the external action
$S_{ext}$, the list of defined fields and external fields.
Moreover the explicit form of the full action is obtained, and 
 the dimension and ghost number of a general integrated
polynomial (in the example $S_{tot}$).
To show the use of the ST operator $s$, an application of
the $Dr$ operator is shown and the variation of the full action
(which turn out to be zero) is reported.
\\
The use of the module \code{Create} is shown and the most general
polynomials of dimension 3 and ghost number -1, and dimension 3 and
ghost number zero are reported.
\\
Finally a simple example showing the nilpotency of the $s$ operator is
reported.

\section{Modules and objects}\label{sec:objects}
In this section we  stretch the most important features of the modules
and objects.
In tab.~\ref{tab:modules} we list the modules with their parameter
with a short description,
in tab.~\ref{tab:operators} and tab.~\ref{tab:over} we give a list of
the operators with their arguments and description.

\begin{table*}{
\caption{List of the modules with their parameter and description}
\begin{tabular}[th]{|l|l|l|}
\hline\label{tab:modules}
Name		&Arguments			&Output\\\hline
Gh  		&Pol      			&Ghost number of \code{Pol}\\
UndefineField	&name, Lorentz			&Delete a Field\\
DefineField	&name, Lorentz, type, dim, gh	&Create a Field\\
UndefineExtField&name, Lorentz			&Delete an ExternalField\\
DefineExtField	&name, Lorentz, type, dim, gh	&Create an ExternalField\\
SetBRST		&name, BRS			&Associate a \code{BRS}
variation to a Field\\
Create		&dimax, gh			&Generate the most
general integrated poly-\\
		&				&nomial of the
fields with dimension less\\
		&				&or equal to \code{dimax}
and ghost number \code{gh}\\
\hline
\end{tabular}
}\end{table*}

\begin{table}{
\caption{List of the operators with their arguments and description}
\begin{tabular}[h]{|l|l|l|}
\hline\label{tab:operators}
Name		&Arguments			&Output\\\hline
Stat		&Mon				&+1 if \code{Mon} is
bosonic, -1 if it's fermionic\\
Dim		&Pol  				&The dimension of \code{Pol}\\
Dr		&Pol, name			&The functional
derivative of \code{Pol} with respect to field \code{name}\\
s		&Pol				&BRST variation of \code{Pol}
\\\hline
\end{tabular}
}\end{table}

\begin{table}{
\caption{List of the overloaded operators with their arguments and description}
\begin{tabular}[h]{|l|l|l|}
\hline\label{tab:over}
Name		&Arguments			&Output\\\hline
NonCommutativeMultiply&Pol1, Pol2		&Take into account the
non commutativity of\\
		      &				&Lie algebra valued
polynomial\\
Wedge		&Pol1, Pol2			&External product
between \code{Pol1} and \code{Pol2}\\
Dot		&Pol1, Pol2			&Scalar product 
between \code{Pol1} and \code{Pol2}
\\\hline
\end{tabular}
}\end{table}

\section*{Acknowledgments}

We would like to thank Prof. R. Ferrari for enlightening conversations
on the subject.
One of us (M.P.) would like to thank Prof. S.P. Sorella for useful
discussion during the preparation of this paper. 
One of us (E.T) would like to thank the hospitality of CERN-TH
division, the department of physics of University of Milan, and
financial support of INFN-CICYT grant.
 
\providecommand{\href}[2]{#2}\begingroup\raggedright\endgroup

\end{document}